\documentclass[%
 reprint,
nofootinbib,
 amsmath,amssymb,
 aps,
 prb,
]{revtex4-2}
\usepackage{graphicx}
\usepackage{dcolumn}
\usepackage{bm}
\usepackage[hidelinks]{hyperref}
\usepackage{color}
\usepackage{physics}
\usepackage{dsfont}
\usepackage{pgf,tikz}
\usepackage{pgfplots}
\pgfplotsset{compat=1.18}
\usepackage{adjustbox}
\usepackage{soul}
\usepackage{amsmath,amssymb,physics}
\usepackage{rotating} 
\usepackage{comment}
\allowdisplaybreaks[3]

\hypersetup{citecolor=red,colorlinks=true,urlcolor=blue}

\usepackage{lmodern}
\usepackage[T1]{fontenc}

\newcommand{\IOPPAS}{Institute of Physics PAS, Aleja Lotnik\'ow 32/46, 02-668 Warszawa, Poland}

\begin{document}

\preprint{APS/123-QED}

\title{
Degeneracy-governed spin squeezing in high-spin Fermi-Hubbard systems weakly coupled to light
}

\author{Hubert Dunikowski and Emilia Witkowska}
\affiliation{\IOPPAS}

\begin{abstract}
High-spin alkaline-earth fermions in optical lattices are promising platforms for spin squeezing beyond the spin-1/2 paradigm.
We show that spin-squeezing dynamics is qualitatively modified by degeneracies inherent to the extended internal spin structure.
We identify these degeneracies as the microscopic origin of the breakdown of the conventional maximal-spin description and develop an effective population-eigenstate framework that quantitatively reproduces the spin-squeezing dynamics. 
Our results establish degeneracy as a generic driver of collective spin dynamics in high-spin systems.
\end{abstract}

\date{\today}

\maketitle

\emph{Introduction---}
Spin squeezing provides one of the most scalable routes toward quantum-enhanced sensing beyond the standard quantum limit~\cite{Kitagawa1993, Wineland1992}, with demonstrated applications in atomic clocks, magnetometers, and interferometers~\cite{Robinson2024,Sewell2014,Cassens2025,Mao2023}.
Fermionic alkaline-earth atoms with large nuclear spin, such as $^{87}$Sr and $^{173}$Yb, are particularly promising platforms for realizing high-spin squeezing~\cite{Cazalilla2014, Gorshkov2010}. Their two-electron singlet ground state decouples the nuclear spin $s$ from the electronic angular momentum, preserving the $\mathrm{SU}(d)$ Hamiltonian symmetry of magnetic sub-levels with $d = 2s+1$, where all spin components interact with equal strength and collisional spin relaxation is strongly suppressed~\cite{Cazalilla2014, Gorshkov2010}.  In optical lattices, these systems realize the high-spin Fermi-Hubbard (FH) model with long coherence times~\cite{Gorshkov2010,Scazza2014} and site-resolved control~\cite{Bloch2008,Taie2012,Hofrichter2016,ferrer2026,plassmann2026}, making them ideal for exploring spin squeezing beyond the spin-1/2 paradigm.

In the strongly repulsive regime with one atom per lattice site, the low-energy physics is governed by the $\mathrm{SU}(d)$ spin-exchange Hamiltonian~\cite{Hermele2009,Manmana2011}. For $s=1/2$ this reduces to the isotropic Heisenberg XXX model, whose $\mathrm{SU}(2)$ symmetry confines the dynamics of any initial spin-coherent state strictly within the maximum-spin Dicke manifold~\cite{Gross2012}. A weak coupling to light breaks the spin-exchange symmetry: virtual spin-wave excitations generate all-to-all spin correlations~\cite{Molmer1999, PhysRevResearch.1.033075, PhysRevResearch.3.013178, Davis2019,PhysRevB.108.104301, PhysRevB.109.214310, Sundar2024}, effectively captured by Lipkin-Meshkov-Glick dynamics~\cite{Lipkin1965}, which reduces to the one-axis twisting (OAT) model~\cite{Kitagawa1993} under periodic boundary conditions~\cite{ Tana2022}. This twisting mechanism is the established route to scalable spin squeezing for $s=1/2$. For $s\ge 1$, however, the $\mathrm{SU}(d)$ symmetry of the spin-exchange model gives rise to an extensive degeneracy that enlarges the zero-energy manifold far beyond the Dicke sector -- a feature with no $s=1/2$ counterpart and, as we show, with profound consequences for spin-squeezing dynamics.

In this work, we investigate spin squeezing in high-spin alkaline-earth FH systems weakly coupled to light in the strongly repulsive regime. Although the structure of Dicke and spin-wave states suggests that the dynamics of an initial spin-coherent state should follow the twisting mechanism established for $s=1/2$, we demonstrate that this picture breaks down for $s\ge 1$. We identify the origin of this breakdown in the extensive eigenstate degeneracy inherent to higher-spin systems, which introduces additional states, characterized here in the basis of population eigenstates that reflect the magnetic sub-level occupation structure~\cite{long}, that coexist within the same energy manifolds as Dicke and spin-wave states. This makes the Dicke projection incomplete and invalidates the collective spin description in the perturbative regime.

Performing the Schrieffer-Wolff (SW) transformation that treats the atom-light coupling perturbatively~\cite{BRAVYI20112793}, we derive corrected effective models that, in contrast to the $s=1/2$ case, exhibit nonconservation of the collective spin, a prediction confirmed by exact many-body simulations, for both coupling schemes considered. For spin-orbit coupling, the corrected effective Hamiltonian contains population-dependent terms that break collective spin conservation, qualitatively modifying the squeezing dynamics. For scalar-tensor coupling, the correction is even more striking: the effective Hamiltonian reduces to a single-body energy shift, incapable of generating the entanglement required for squeezing. These results establish degeneracy as an intrinsic and previously overlooked factor governing collective spin dynamics in any high-spin system coupled to light.

\begin{figure}[]
    \includegraphics[width=\linewidth]{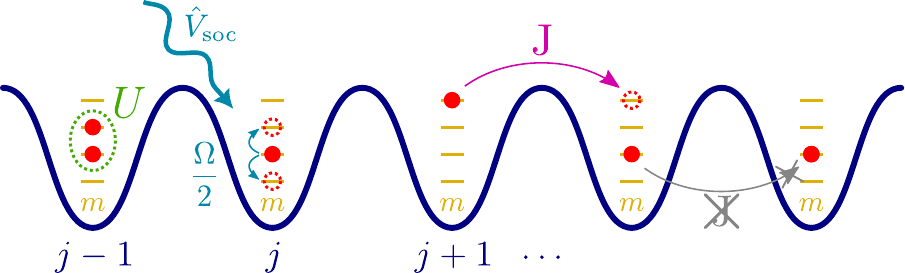}
    \caption{Schematic of the processes considered with FH Hamiltonian, highlighting on-site interactions $U$, tunnelling $\mathrm{J}$ between neighboring atoms and coupling $\hat{V}_{\rm soc}$ with light. Multiple atoms may occupy a lattice site $j$ only at different magnetic levels $m$, as required by Pauli exclusion.}
    \label{fig:fig1}
\end{figure}

\emph{The model---}
We consider $N$ fermionic alkaline-earth atoms of spin $s$ in a one-dimensional optical lattice~\cite{Gorshkov2010}. 
Each of $j=1,\ldots,N$ lattice sites hosts $d=2s+1$ internal states labeled by magnetic quantum number $m=-s,\ldots,s$, with creation operators $\hat{a}_{j,m}^\dagger$. The total atom number $N=\sum_j \hat{n}_j$ is conserved, where $\hat{n}_j=\sum_m \hat{a}_{j,m}^\dagger \hat{a}_{j,m}$. We assume periodic boundary conditions. The system is described by the Fermi-Hubbard Hamiltonian,
\begin{align}
\label{eq:FHm}
    \hat{H}_{\mathrm{FH}} &= \hat{H}_{\mathrm{t}} + \hat{H}_{\mathrm{int}} ,\\
    \label{eq:FHtunneling}
    \hat{H}_{\mathrm{t}} &= -\mathrm{J}\sum\limits_{j}\sum\limits_{m=-s}^{s} (\hat a_{j+1,m}^\dagger \hat a_{j,m} + \text{h.c.}) ,\\
    \label{eq:FHint}
    \hat{H}_{\mathrm{int }} &= \frac{U}{2}\sum\limits_{j=1}^{N} \hat n_j(\hat n_j-1)\, ,
\end{align}
containing tunneling process \eqref{eq:FHtunneling} between neighboring sites governed by $\mathrm{J}$ and on-site repulsive interaction term~\eqref{eq:FHint} with strength $U$, as illustrated in Fig.~\ref{fig:fig1}.

In the strongly repulsive regime, when $U\gg\mathrm{J}$, with one atom in each lattice site second-order virtual tunneling processes map the FH Hamiltonian \eqref{eq:FHm} onto an effective nearest-neighbor spin-exchange model~\cite{Gorshkov2010,Cazalilla2014}, 
\begin{equation}
    \hat{H}_\mathrm{SE} = \mathrm{J_{SE}}\sum\limits_{j}
\,\sum\limits_{m\neq m'}\!\! \Big(\hat S^{m'\rightarrow m}_j\, \hat S^{m\rightarrow m'}_{j +1} {-}\hat S^{m\rightarrow m}_j\, \hat S^{m'\rightarrow m'}_{j+1} \Big),
\label{eq:SEM}
\end{equation}
with $\mathrm{J_{SE}}=2\mathrm{J^2}/U$ and $\hat S^{m'\rightarrow m}_j=\hat a_{j,m}^\dagger \hat a_{j,m'}$ referred to as S-arrow operators.

The local spin rising and lowering operators are defined in terms of S-arrow operators as 
$\hat{J}_{+,j} = \sum_{m=-s}^{s-1}\! \alpha_{s,m}
\hat{S}_j^{m\rightarrow m+1}$ and
$\hat{J}_{-,j} = \sum_{m=-s}^{s-1}\! \alpha_{s,m}\ \hat{S}_j^{m+1\rightarrow m}$,
where $\alpha_{s,m} = \sqrt{(s-m)(s+m+1)}$.
The local spin components $\hat{J}_{\sigma}$ obeying cyclic commutation relations $[\hat{J}_{\sigma},\hat{J}_{\sigma'}]= i \epsilon_{\sigma \sigma'\sigma''}\,\hat{J}_{\sigma''}$
are 
$\hat{J}_{x,j }  = (\hat{J}_{+,j}+\hat{J}_{-,j})/2$, $\hat{J}_{y,j }  = (\hat{J}_{+,j}-\hat{J}_{-,j})/(2\mathrm i)$ and 
$\hat{J}_{z,j} = \sum_{m=-s}^{s}m \, \hat n_m$, where $\hat n_m=\sum_j \hat a_{j, m}^\dagger \hat a_{j, m}$ are total populations of each magnetic level $m$.
The collective spin operators are
$\hat{J}_\sigma = \sum_j \hat{J}_{\sigma,j}$.

For $s\ge 1$, the S-arrow algebra is not equivalent with the spin-$s$ representation $\hat{J}_{\sigma,j}$, and thus local spin operators are not identical to S-arrow operators. This identification is valid only for $s=1/2$ where the XXX Heisenberg spin model is realized.

The SE Hamiltonian \eqref{eq:SEM} conserves both the total collective spin $\hat{J}^2=\sum_{\sigma}\hat{J}_\sigma^2$ and its $z$-projection $\hat{J}_z$, and therefore admits a block-diagonal decomposition in their eigenbasis. We concentrate on two eigenstate energy manifolds that are directly connected by the atom-light coupling: the zero-energy manifold $E=0$, relevant for the unperturbed initial state dynamics, and the spin-wave manifold at energy $E_q$, which enters through virtual excitations in the perturbative treatment.

The first manifold has energy $E=0$ and includes the maximal total spin $\Lambda=sN$ Dicke states: $|\Lambda,M\rangle \propto \hat{J}_-^{\Lambda-M} |\mathrm{max}\rangle$, where $|\mathrm{max}\rangle = \bigotimes_{j=1}^N |s\rangle_j$ and $|s\rangle_j$ denotes the single-atom in site $j$ in the magnetic level $m=s$. The second manifold has energy $E_q = 2\mathrm{J_{SE}}(\cos q\lambda-1)$ and includes the spin-wave states
$|q\rangle \propto \sum_j e^{i q j \lambda}\, \hat{J}_{j,-} |\Lambda,M\rangle$, where $q = \frac{2\pi k}{N\lambda}$ with $k \in \{0,1,\ldots,N-1\}$ and $\lambda$ is the lattice spacing. Both the Dicke and the spin-wave states are eigenstates of the SE Hamiltonian \eqref{eq:SEM} and coincide with those of the isotropic Heisenberg XXX model.

In what follows, we concentrate on the total spin dynamics $\hat{J}^2$ and the evolution of the Wineland spin squeezing parameter $\xi^2 = 2 s N(\Delta J_{\perp,\mathrm{min}})^2/|\langle \hat{J}\rangle|^2$~\cite{Wineland1992}, where $(\Delta J_{\perp,\mathrm{min}})^2$ denotes the minimal fluctuations of the collective spin transverse to the mean-spin direction $\langle \hat{J}\rangle$ for $N$ spin-$s$ particles.
We consider the initial spin coherent state  $|\vartheta,\varphi\rangle = e^{-\mathrm{i}\hat{J}_z \varphi} e^{-\mathrm{i} \hat{J}_y \vartheta}|\mathrm{max}\rangle$, here taken at $\vartheta=\pi/2$, $\varphi=0$, where squeezing develops from quantum fluctuations of a fully polarized state.

\emph{Coupling to light---} The SE Hamiltonian \eqref{eq:SEM} does not generate nontrivial spin dynamics from any initial spin-coherent state $|\pi/2,0\rangle $. A coupling to light introduces an additional process $\hat V$ to the system Hamiltonian, with strength small compared to the spin-exchange energy scale, therefore is can be treated within the SW expansion~\cite{BRAVYI20112793}. Up to second order, the effective Hamiltonian describing dynamics within the $E=0$ energy sector is a sum of the two terms
\begin{equation}
    \hat{H}_{\rm eff}^{(0)} = \hat{I}_0 V \hat{I}_0, 
    \qquad
    \hat{H}_{\rm eff}^{(2)} = \hat{I}_0 V \hat{G}_q V \hat{I}_0,
    \label{eq:HSWgeneral0}
\end{equation}
where $\hat{I}_0$ and $\hat{G}_q$ operate in $E=0$ and $E=E_q$ energy manifolds of \eqref{eq:SEM}.

\begin{figure}[]
    \centering
    \includegraphics[width=1.0\linewidth]{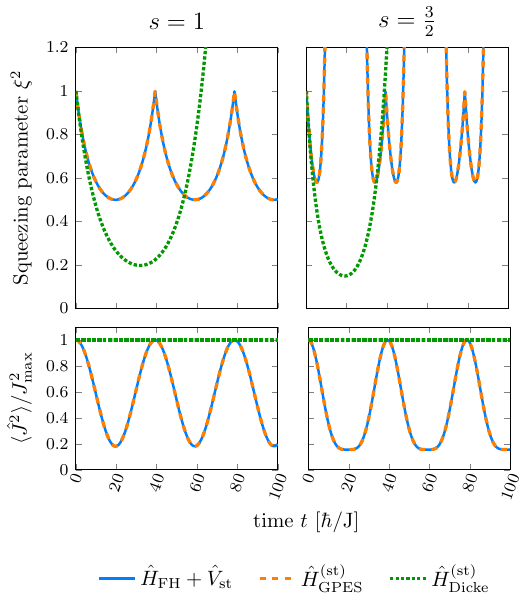}
    \caption{
    Temporal variation of spin squeezing $\xi^2$ (top panels) and total spin $\langle \hat J^2\rangle$ (bottom panels) of the system with a weak scalar-tensor coupling ($\hat V_{\rm st}$) for $s=1$ (left panels),  $s=3/2$ (right panels), and  
    $N=10$, $U=4 \mathrm{J}$ and $\Omega_{\mathrm{st}}=0.08 \mathrm{J}$, where $J^2_{\mathrm{max}}=\hbar^2sN(sN+1)$. The solid blue and dashed orange lines overlap.
    }
    \label{fig:st}
\end{figure}

As a first example, a linearly polarized beam perpendicular to the lattice generates scalar and tensor light shifts~\cite{Domantas2024}, 
\begin{equation}
    \hat{V}_{\rm st}=\Omega_{\rm st} \sum_{j=1}^N \left(\hat{J}_{z,j}^2 - \frac{1}{3}\hat{J}_{j}^2 \right).
    \label{eq:2ndcoupling}
\end{equation}
Projecting onto the Dicke manifold $\hat{I}_0=\sum_M |\Lambda,M\rangle\langle \Lambda,M|$ in the leading zero order yields~\cite{NavidJakub}
\begin{equation}
    \hat{H}_{\rm Dicke}^{(\mathrm{st})}=\frac{\Omega_{\rm st}(2s-1)}{2Ns-1}\hat{J}_z^2,
    \label{eq:heff-st-dick}
\end{equation}
where we omitted constant energy terms.
This is the well-known analytically solvable OAT model~\cite{Kitagawa1993, PhysRevB.109.214310}, which conserves the total spin $\hat{J}^2$ achieving the optimal scaling of the Wineland parameter $\xi_{\mathrm{best}}^2 \simeq \Lambda^{-2/3}$ at the optimal evolution time. Comparison with the exact FH solution reveals a clear discrepancy: while the OAT model predicts smooth squeezing with conserved total spin, the exact dynamics shows irregular oscillations of $\xi^2$ accompanied by strong variations of $\langle \hat{J}^2 \rangle $ well below its initial value (Fig.~\ref{fig:st}), unambiguously signaling a breakdown of the Dicke projection.

\begin{figure}[]
    \centering
    \includegraphics[width=1.0\linewidth]{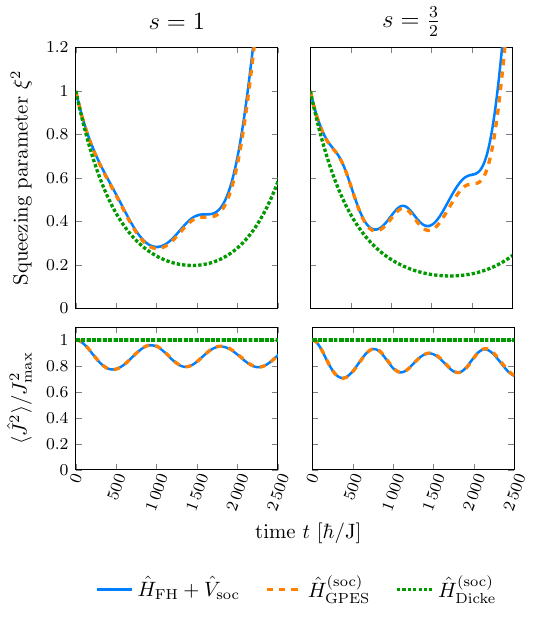}
    \caption{
    Evolution of spin squeezing parameter $\xi^2$ and the total spin $\langle \hat J^2\rangle$ a weak spin-orbit coupling ($\hat V_{\rm soc}$) for $\Omega_{\mathrm{soc}}=0.08 \mathrm{J}$, $\phi=4\cdot\frac{2\pi}{N}$ and other parameters as in Fig.~\ref{fig:st}.
    }
    \label{fig:soc}
\end{figure}

As a second example, a spin-orbit-type coupling is considered~\cite{Domantas2024},
\begin{equation}
    \hat{V}_{\rm soc}=\frac{\hbar\Omega_{\rm soc}}{2}\sum_{j=1}^N \left(e^{i\phi j}\hat{J}_{j,+}+e^{-i\phi j}\hat{J}_{j,-}\right).
    \label{eq:soc}
\end{equation}
The SW projection, up to the second order including the Dicke manifold $\hat{I}_0=\sum_M |\Lambda,M\rangle\langle \Lambda,M|$ and virtual spin-wave processes via $\hat{G}_q=\sum_{q\neq 0}\frac{|q\rangle\langle q|}{-E_q}$, leads to OAT
\begin{equation}
    \hat{H}_{\rm Dicke}^{(\mathrm{soc})}
    =\frac{\hbar^2\Omega_{\rm soc}^2}{4\mathrm{J}_{\rm SE}(1-\cos\phi)}\,
    \frac{1}{2Ns-1}\,\hat{J}_z^2,
    \label{eq:effJzdicke}
\end{equation}
for $\phi\neq\pi$. As in the previous example, this effective description fails: the exact dynamics exhibits qualitative deviations in both $\xi^2$ and $\langle \hat J^2 \rangle$ from the OAT prediction, with the total spin showing pronounced nonconservation (Fig.~\ref{fig:soc}).

While this construction is valid for $s=1/2$~\cite{Tana2022}, for $s\ge 1$ the Dicke projection is structurally incomplete regardless of the values of $J$ and $U$, since additional degenerate states at both $E=0$ and $E=E_q$ emerge from the enlarged internal spin structure and are not accounted for in the SW projection~\eqref{eq:HSWgeneral0}. A correct description requires extending the SW projection to the full degenerate manifold, as developed in the next section.

\emph{Missing eigenstates of the SE Hamiltonian---}
To construct eigenstates of \eqref{eq:SEM} in the $E=0$ and $E=E_q$ manifolds, it is convenient to work in the basis of population eigenstates (PES), specified by the occupation vector $\vec{n}=(n_s,n_{s-1},\ldots,n_{-s})$, where $n_m$ denotes the total population in each magnetic level. Physically, PES are many-body Fock states in the magnetic level occupation. Unlike Dicke states, which are coherent superpositions carrying maximal collective spin, PES carry no a priori collective spin character and span a much larger portion of the Hilbert space.

The zero-energy manifold is spanned by population eigenstates, referred to as GPES,
\begin{equation}
    |\vec{n} \rangle 
    = \mathcal{N}_{\vec{n}}^{-1/2}
    \prod_{m=-s}^{s-1} \left( \hat{S}^{s \rightarrow m} \right)^{n_m} |\text{max}\rangle 
\end{equation}
with normalization $ \mathcal{N}_{\vec{n}} = N! \ n_{-s}!\ n_{-s+1}! \ldots n_{s-1}!/ n_s! $
, see~\cite{long} for details.
These states are mutually orthogonal, $\langle \vec{n}|\vec{n}'\rangle=\delta_{\vec{n},\vec{n}'}$. 

Only specific GPES, namely $|\mathrm{max}\rangle$ and $|\mathrm{min}\rangle=\bigotimes_{j=1}^N |{-}s\rangle_j$, corresponds to a Dicke state with $\Lambda=sN$, $M=sN$ and $M=-sN$. All other Dicke states arise as specific linear combinations of GPES, but they do not span the whole GPES manifold. The dramatic difference in the size of the two spaces, $ \#\{\mathrm{Dicke}\} = 2sN + 1 $ versus $
\#\{\mathrm{GPES}\}=\frac{(N+2s)!}{N!(2s)!}$ with equality only for $s=1/2$, quantifies the extent of this incompleteness. 
We have $ \#\{\mathrm{Dicke}\} = 91 $ and $
\#\{\mathrm{GPES}\}=92\,378$ for ${}^{87}$Sr with $s=9/2$ and $N=10$, illustrating how severely the Dicke sector underrepresents the full zero-energy manifold. 

The second manifold of population eigenstates consists of states $|q,m_q;\vec{n}'\rangle$ corresponding to the SWS energy $E_q= 2\mathrm{J_{SE}}\big(\cos qd -1\big)$, hereafter referred to as $q$-PES
\begin{equation}
    |q,m_q;\vec{n}'\rangle = (\mathcal{Q}_{n_{m_q}}^{n_{m'}})^{-1/2}\,\hat{S}^{m'\rightarrow m_q}_q\,|\vec{n}\rangle\label{eq:q-PES}
\end{equation}
where $\hat{S}^{m'\rightarrow m_q}_q = \sum_{j=1}^N e^{\mathrm{i}q\lambda j}\, \hat{S}^{m'\rightarrow m_q}_j$, 
and $\vec{n}'=(n_s,\ldots,n_{m'}-1,\ldots,n_{m_q}\!+1,\ldots,n_{-s})$.
The normalization factor is $\mathcal{Q}_{n_{m_q}}^{n_{m'}}=n_{m'}(N-n_{m_q}\!-1)/(N-1)$. The $q$-PES are orthogonal in $\vec{n}$ and $q$, $\langle q,m_q; \vec{n}| q', m_{q}; \vec{n}' \rangle = \delta_{q,q'}\,\delta_{\vec{n},\vec{n}'}$. 

The two specific $q$-PES, namely $|q, m_q = s-1, \vec{n} = (N-1,1,0,\ldots)\rangle$ and $|q,m_q{=}{-}s{+}1;\vec{n}{=}(0,\ldots,0,1,N{-}1)\rangle$ corresponds to SWS with $\Lambda = sN - 1$, $M = sN - 1$ and $M = -sN + 1$.
The conventional spin-wave states are embedded within the broader $q$-PES manifold as special cases, precisely as Dicke states are embedded within GPES. The vast majority of $q$-PES carry not only $\Lambda=sN-1$ collective spin but also $\Lambda < sN-1$ and have no counterpart in the standard spin-wave description, confirming that the $E_q$ energy manifold is equally affected by the degeneracy-driven enlargement of the Hilbert space.
The number of independent spin-wave states for a given $q$ is $\#\{\mathrm{SWS}\}=2sN-1$, whereas the number of $q$-PES is
$\#\{q\text{-PES}\}=\sum\limits_{k=2}^{d}d!\, (N-1)!/[{(k-2)!\, k!\,(N-k)!\,(d-k)!}]
$ (where $d=2s+1$). 
Taking ${}^{87}$Sr with $s=9/2$ and $N=10$ as an example, we have $\#\{\mathrm{SWS}\}=89$ and $
\#\{q\text{-PES}\}=3\,544\,398$.

The coexistence of GPES and Dicke states within the same $E=0$ energy manifold has direct consequences for the SW projection: the light coupling connects Dicke states not only to spin-wave states but also to the broader $q$-PES manifold. In the reverse process, it couples all spin-wave states with all the GPES states, not only to Dicke states. A SW projection restricted to the Dicke sector therefore misses these additional matrix elements entirely, yielding an incorrect effective model.

One can show that both $E=0$ and $E=E_q$ degenerate eigenstates can be represented in terms of total-spin states $|\Lambda,M\rangle$ including the maximal $\Lambda=s N$ and lower $\Lambda<sN$ total spin states, see~\cite{long} for details. This incomplete representation of spin states for $s\ge1$ underlies the failure of the previously obtained effective collective-spin models.

\emph{Corrected effective models for collective observables---}
Including the full set of GPES and $q$-PES eigenstates of the SE Hamiltonian \eqref{eq:SEM} allows one to go beyond the Dicke and SW states. Accordingly, the low-energy projector is taken over the complete zero-energy manifold spanned by GPES,
$\hat{I}_0=\sum_{\vec n} |\vec n\rangle\langle \vec n|$,
while the virtual processes involving $E_q$ energy states contribute in 
$\hat G_q=\sum_{q\ne 0}\frac{\hat{I}(q\text{-PES}) }{-E_q}$, where $\hat{I}(q\text{-PES})$ is projector on the $q$-PES states, see also ~\cite{long} for details.

In the GPES basis, the scalar-tensor coupling \eqref{eq:2ndcoupling} acts solely on the total population of each magnetic sub-level $n_m$, without generating coherences or correlations between different levels. The projection onto the full GPES manifold therefore averages out the two-body interaction structure present in the Dicke projection, leaving only a single-body energy shift that commutes with all population operators and generates no entanglement. Therefore, we have
\begin{equation}
    \hat{H}_{\rm GPES}^{(\mathrm{st})}=\Omega_\mathrm{st} \sum_{m=-s}^s m^2 \hat{N}_m. 
    \label{eq:Heff-st-cor}
\end{equation}
The corrected model accurately reproduces the exact FH dynamics and  Fig.~\ref{fig:st} confirms this. While the Dicke projection predicts smooth OAT-like squeezing with conserved total spin, the corrected GPES model accurately reproduces both the suppression of squeezing and the pronounced oscillatory variations of $\langle \hat{J}^2\rangle$ observed in the exact FH dynamics, with the solid blue and dashed orange lines overlapping throughout the perturbative regime $\Omega_{\rm st} \ll {\rm J}_{\rm SE}$.

For the spin-orbit coupling \eqref{eq:soc} with $\phi\equiv q\lambda \neq\pi$, the effective Hamiltonian becomes
\begin{equation}
     \hat{H}^{(\mathrm{soc},\phi)}_{\rm GPES}=\chi_\phi
     \left[
     \hat J^2 - \hat{J}_z^2
     -
     \frac{N}{2}\sum_{m=-s}^{s-1}
     \alpha_{s,m}^2
     (\hat n_m+\hat n_{m+1})
     \right],
    \label{eq:corr1}
\end{equation}
with
\begin{equation}
    \chi_\phi = \frac{\hbar^2\Omega_{\rm soc}^2}
     {4\mathrm{J_{SE}}(\cos \phi-1)(N-1)},
\end{equation}
where the population-dependent term breaks conservation of the total spin. For $s=1/2$, however, this contribution reduces to the conserved total atom number, recovering the OAT model with emergent $\hat J_z^2$ dynamics as in~\cite{Tana2022}.

In the case $\phi=\pi$, one obtains
\begin{multline}
\hat{H}^{(\mathrm{soc},\pi)}_{\text{GPES}}=
    \chi_\pi
    \left[    
    4\hat{J}_x^2-  N\!\sum\limits_{m=-s}^{s-1}  \alpha_{s,m}^2\Big(\hat{n}_m+\hat{n}_{m+1}\Big) \right.
    \label{eq:corr2}\\
\left. -  N\!\sum\limits_{m=-s}^{s-2}  \alpha_{s,m}\alpha_{s,m+1}\Big(\hat{S}^{m\rightarrow m+2} + \hat{S}^{m+2\rightarrow m}\Big)
    \right]
\end{multline}
with
\begin{equation}
    \chi_\pi=-\frac{\hbar^2\Omega_{\rm soc}^2}{8\mathrm{J_{SE}}(N-1)}.
\end{equation}
Again, for $s=1/2$ the last term vanishes identically, while the second reduces to a constant, yielding an effective OAT Hamiltonian proportional to $\hat J_x^2$. 
Example evolution of spin squeezing in Fig.~\ref{fig:soc} confirms that the GPES model Eq.~\eqref{eq:corr1} tracks the exact FH solution closely, while the Dicke projection fails qualitatively, predicting smooth OAT squeezing with conserved total spin, in stark contrast to the irregular oscillations and pronounced nonconservation of total spin observed in the exact dynamics. The discrepancy grows with increasing spin $s$, consistent with the combinatorial growth of the GPES manifold relative to the Dicke sector.

The terms proportional to the magnetic sub-level populations and S-arrow operators in the effective models~\eqref{eq:corr1} and \eqref{eq:corr2} are a direct consequence of the degeneracy of high-spin systems. 
Both the twisting term and the population-dependent term scale as $N^2$ in the thermodynamic limit confirming that degeneracy-induced terms are not a finite-size artifact but an intrinsic feature of the effective dynamics that persists in the thermodynamic limit

The derivation of the two effective Hamiltonians, \eqref{eq:corr1} and \eqref{eq:corr2}, relies on a two-step perturbative treatment: the strongly repulsive regime $\mathrm{J}\ll U$ and the weak atom--light coupling $\Omega_{\mathrm{soc}}\ll {\rm J}_{\rm SE}$. The validity of the effective models requires both perturbative conditions to be satisfied simultaneously; a detailed benchmarking against exact many-body simulations and the ranges of validity are discussed in Appendix~\ref{app:AppendixB}.

\emph{Conclusions---}
We have shown that light-induced spin-squeezing dynamics in high-spin alkaline-earth FH systems is governed by the extensive degeneracy of the spin-exchange spectrum, a direct consequence of the multilevel internal structure absent in the spin-1/2 case.
The key insight is that for $s\ge 1$, the $E=0$ and $E=E_q$ energy manifolds are vastly larger than their Dicke and spin-wave counterparts, and any realistic light coupling inevitably activates this broader class of states, making the Dicke description structurally incomplete.
These results are directly relevant to ongoing experiments with $^{87}$Sr and $^{173}$Yb in optical lattices~\cite{ferrer2026,plassmann2026}, where site-resolved imaging and spin-dependent probes make possible to observe the degeneracy-governed spin-squeezing dynamics predicted here directly.

\emph{Acknowledgments---}
The Authors gratefully acknowledge discussions with B. Laburthe-Tolra and M. Robert-de-Saint-Vincent. 
E.W. acknowledges the kind hospitality of the Laboratoire de Physique des Lasers.
This work was supported by the Polish National Science Center SHENG project DEC-2023/48/Q/ST2/00087.

\appendix

\section{Range of validity of the effective models}
\label{app:AppendixB}

\begin{figure*}[]
    \centering
    \includegraphics[width=1.0\textwidth]{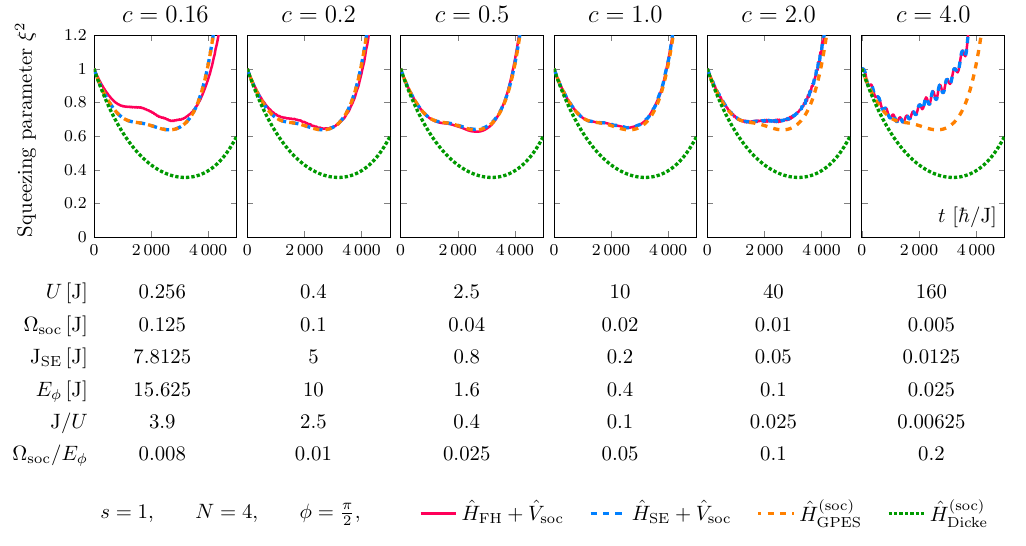}
    \caption{
    Examples of the four dynamical descriptions considered in this work,
$\hat{H}_{\mathrm{FH}}+\hat{V}_{\mathrm{soc}}$,
$\hat{H}_{\mathrm{SE}}+\hat{V}_{\mathrm{soc}}$,
$\hat{H}_{\mathrm{GPES}}^{(\mathrm{soc})}$,
and
$\hat{H}_{\mathrm{Dicke}}^{(\mathrm{soc})}$,
shown on a common timescale determined by $\chi=1/12\,000$, for different values of $(U,\Omega_{\mathrm{soc}})$.
The dynamics generated by $\hat{H}_{\mathrm{FH}}+\hat{V}_{\mathrm{soc}}$ (solid red line) serve as the most accurate reference.
The spin-exchange description $\hat{H}_{\mathrm{SE}}+\hat{V}_{\mathrm{soc}}$ remains quantitatively accurate for $\mathrm{J}/U\lesssim 2.0$, while the effective GPES Hamiltonian $\hat{H}_{\mathrm{GPES}}^{(\mathrm{soc})}$ accurately reproduces the dynamics for $\Omega_{\mathrm{soc}}/E_\phi\lesssim 0.05$.
Within the range of system sizes and atomic spins investigated, the bounds $\mathrm{J}/U\lesssim 2.0$ and $\Omega_{\mathrm{soc}}/E_\phi\lesssim 0.05$ appear to be largely universal.
}
    \label{fig:squeez}
\end{figure*}

The characteristic timescale of the effective dynamics is determined by the coupling constant $\chi$. For fixed $N$ and $\phi$, it scales as $\chi\propto\frac{\Omega_{\mathrm{soc}}^2 U}{\mathrm{J}^2}$. Choosing $\mathrm{J}$ as the unit of energy, one finds that the transformation $U\rightarrow c^2U$ and $\Omega_{\mathrm{soc}}\to \Omega_{\mathrm{soc}}/c$ leaves $\chi$ unchanged. Therefore, different parameter sets related by this transformation generate dynamics on the same effective timescale while probing different depths of the two-step perturbative regimes.

Careful scaling analysis shows that increasing $c$ improves the validity of the spin-exchange approximation, as we have $\mathrm{J}/U\to c^{-2}\mathrm{J}/U$, and weakens the validity of the perturbative treatment of the light coupling, because
$\Omega_{\mathrm{soc}}/{\rm J}_{\rm SE} \rightarrow
c\, \Omega_{\mathrm{soc}}/{\rm J}_{\rm SE}$ from $\mathrm{J}_{\rm SE}\rightarrow \mathrm{J}_{\rm SE}/c^2$. Decreasing $c$ has the opposite effect: it improves the light-coupling perturbative regime while reducing the accuracy of the spin-exchange approximation. 
As a result, for a given reference parameter set there exists only a finite range of $c$ values for which both perturbative descriptions remain simultaneously valid.

To quantify these limits, we benchmark the effective theories against exact many-body simulations. Fig.~\ref{fig:squeez} compares the dynamics obtained from the exact $\hat H_{\mathrm{FH}}+\hat V_{\mathrm{soc}}$ Hamiltonian with those generated by the effective models \eqref{eq:corr1} and \eqref{eq:effJzdicke} at a fixed timescale $\chi=1/12\,000$ for several parameter sets related by the above scaling transformation. The table below summarizes the parameter values used in the calculations shown in the respective panels of Fig.~\ref{fig:squeez}.
From these observations, we find that the perturbative treatment of the light coupling remains quantitatively accurate for $\Omega_{\mathrm{soc}}/E_{q=\phi}\lesssim0.05$, indicating that a relatively deep perturbative regime is required for the effective model to faithfully reproduce the exact squeezing dynamics. In contrast, the spin-exchange description remains remarkably robust, providing accurate results up to $\mathrm{J}/U\lesssim2.0$, well beyond the nominal strong-coupling regime in which it is derived. This robustness suggests that the effective Hamiltonian $\hat{H}_{\mathrm{GPES}}^{(\mathrm{soc})}$ captures the essential squeezing dynamics even outside the deep Mott-insulating regime.

Within the range of system sizes and atomic spins investigated, the bounds $\Omega_{\mathrm{soc}}/E_{q=\phi}\lesssim0.05$ and $\mathrm{J}/U\lesssim2.0$ appear to be largely insensitive to both $N$ and $s$.

\bibliography{biblio}

\end{document}